\documentclass[aps,prl,twocolumn,tightenlines,showpacs,nofootinbib]{revtex4}
\usepackage{bm}
\usepackage{graphics}
\usepackage{epsfig}
\usepackage{graphicx}
\usepackage{amsmath}
\usepackage{amsfonts}
\usepackage{amssymb}
\begin{document}

\title{Phase Transitions in Parallel Replication Process}
\author{P. N. Timonin}
\email{timonin@aaanet.ru}
\affiliation{Physics Research Institute at
Southern Federal University, 344090 Rostov-on-Don, Russia}
\author{G. Y. Chitov }
\email{gchitov@laurentian.ca}
\affiliation{Department of Physics, Laurentian University,
Sudbury, Ontario P3E 2C6, Canada}

\date{\today}

\begin{abstract}
The one-dimensional kinetic contact process with parallel update is introduced
and studied by Monte Carlo simulations. This process is proposed to describe
the plant population replication and epidemic disease spreading among them. The
phase diagram of the model features the line of the second order transitions
between absorbing and active phases. The numerical results for the critical
index $\beta$ demonstrate its continuous variation along the transition line
accompanied by the variations of the structural characteristics of limiting
steady states. We conjecture the non-universality of the critical behavior of
the model.
\end{abstract}


\maketitle

There has been a steadily growing interest during the last two decades or so in
the kinetics and phase transitions in non-equilibrium systems. The applications
of those systems range from physics, like, e.g., statistical physics, critical
phenomena, condensed matter to less conventional fields, like biology, ecology
or quantitative finance. \cite{Hinrichsen00,Odor04,Ferreira04} The essentially
non-equilibrium feature of these systems is due to their possibility to
irreversibly enter an absorbing (dead, empty) state. The mere existence of such
a state violates the detailed balance. The central problem in studies of
non-equilibrium systems is the transitions they undergo between various active
phases and the inactive (absorbing) state.

The kinetic contact processes which model, e.g., the epidemic disease spreading
or population replication, are the simplest kinetic models exhibiting such
non-equilibrium phase transitions under variation of their parameters.
\cite{Hinrichsen00,Odor04,Ferreira04} These variations consist in the change of
limiting probability distribution for the site occupation numbers in
replication process, or the infected site numbers for epidemic models. Mostly
these models possess two phases: the absorbing state with the population or
viruses extinct and the active phase where some sites are populated or
infected.

Usually the sequential update formalism is used to study these models. It is
based on the differential kinetic equation for the probability distribution
function. In this framework the generic second order transition into absorbing
state of the directed percolation (DP) universality class is established for
the majority of such contact processes. \cite{Hinrichsen00} Another approach is
the parallel update scheme when the model is represented as a discrete-time
probabilistic cellular automaton (PCA). \cite{Kinzel85,Rujan87,Georges89} It is
known that these two approaches can give rather different results
\cite{Hinrichsen00} and one may reasonably suppose that in such case parallel
update gives more adequate description of real kinetics just because in Nature
there is no queue for the fulfillment of the state transformation processes. So
the studies of PCA implementation of replication (epidemic) processes may give
a more realistic picture of them and reveal some new types of critical
behavior.

In this paper we consider a one-dimensional PCA representing the contact
process of population replication or disease spreading. Imagine a line of
plants which maintain population by spreading the seeds to the nearby sites so
the new plants can grow on them if these sites are empty. Let $q$ be the
probability of such event for empty site having only one neighboring plant. For
the case of two plants around the empty one we can choose this probability $r >
q$ to be $r = 1- (1-q)^2 = q(2-q)$  assuming the independence of the two
plants' seeds ``not spreading effects''. Some other choices are also possible
but here we consider only the case $r = q(2-q)$. Let also $p$ be the
probability for plant to survive in the one time step. Thus we have the PCA on
the line of two-state sites (empty and full $\mapsto~S_i = 0, 1$ resp., where
$i$ is the intrachain site index) with the evolution step probabilities defined
for the local three-site configurations in Table \ref{tab:table1}.

\begin{table}[h]
\caption{\label{tab:table1} Automaton rules }
\begin{ruledtabular}
\begin{tabular}{ccccc}
$S_{i-1}, S_i, S_{i+1}$ &111, 110, 011, 010~ & 100, 001 & 101 & 000 \\
Probability ($S_i =1$) & $p$ & $q$ & $q(2-q)$ & 0 \\
\end{tabular}
\end{ruledtabular}
\end{table}
Obviously, this model describes also the epidemic process among the plants,
since instead of the plant seeds we can consider the virus spreading. Note also
that this process differs essentially from the one of the cell replication
considered in Ref.[\onlinecite{Ferreira04}].

\begin{figure}[htp]
\centering
\includegraphics[height=6.0cm]{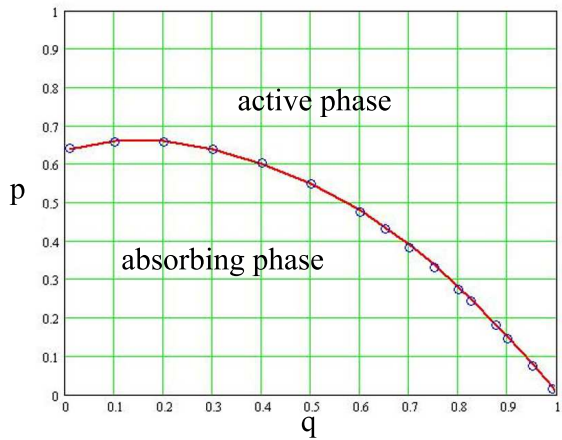}
\caption{(Color online) Phase diagram.}\label{Fig.1}
\end{figure}

We have performed Monte Carlo simulations of this PCA starting from several
random initial states for the chains of $N$  = 10000 sites for up to 5000 time
steps. The periodic boundary conditions were implemented. The resulting $p-q$
diagram (Fig. \ref{Fig.1}) consists of a single transition line between
absorbing (all occupation numbers $S_i = 0$ in the infinite-time limit) and
active (some $S_i \neq 0$) phases. The curve has the following approximate
analytical representation $q = 0.66-0.913(p - 0.15)^2$, determined from a
direct fit.

\begin{figure}[htp]
\centering
\includegraphics[height=6.0cm]{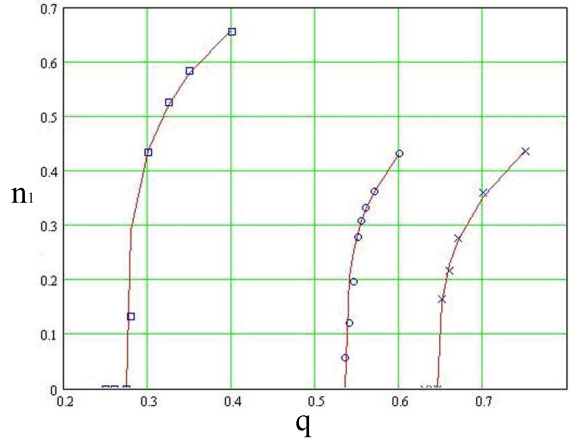}
\caption{(Color online) $q$-dependence of $n_1$ for $p$ = 0.2 ($\times$), 0.5
($\circ$), 0.8 ($\square$). Solid lines show power law dependencies with the
indices given in Table  \ref{tab:table2}.}\label{Fig.2}
\end{figure}

We assume the second order transition between the phases, since the
concentration of active sites defined as
\begin{equation}
\label{n1}
    n_1  \equiv  \frac{1}{N} \sum\limits_{i = 1}^N S_i
\end{equation}
appears to be continuous. The examples of such behavior are presented in Fig.
\ref{Fig.2}, where the (continuous) $q$-dependence of $n_1$ is shown for $p =
0.2, 0.5, 0.8$. We found that the points obtained in simulations follow closely
the power law $n_1 \propto (q-q_c)^\beta$ with the indices $\beta$ given in
Table \ref{tab:table2}.

\begin{table}[h]
\caption{\label{tab:table2}
Transition points $q_c$ and critical indices $\beta$ for different $p$ values.  }
\begin{ruledtabular}
\begin{tabular}{cccc}
$p$ & 0.2 & 0.5 & 0.8 \\
$q_c$ & 0.645 & 0.535 & 0.275 \\
$\beta$ & 0.332 & 0.286 & 0.254 \\
\end{tabular}
\end{ruledtabular}
\end{table}

One can notice the unusual variation the index $\beta$ with $q$ instead of its
more conventional universality. The index varies closely to the directed
percolation value $\beta_{DP} = 0.276$. \cite{Hinrichsen00,Odor04} A naive
explanation of these deviations could have been as an artefact of low data
precision, while the PCA studied belongs actually to the DP class. Indeed,
there are small variations of $n_1$ for different trials (and initial states)
as well as its fluctuations in the nominally steady state at large times. Not
too close to the transition point ($\left| {q - q_c } \right| \gtrsim 0.05$)
these variations of $n_1$ are of the order of several percents and could not
possibly account for the differences in $\beta$ of the order of ten percents.

Therefore we are lead to the assumption that this replication process does not
belong to the DP universality class, and it is characterized by the variable
critical indices. To date the signs of the non-universal behavior are found
only in the $1d$ pair contact process with diffusion \cite{Noh04} but the type
of its critical anomalies is still debated; see Ref.
[\onlinecite{Hinrichsen06}] and references therein.  The non-universality of
some equilibrium spin models (i.e., the coupling-dependent critical exponents)
is well known. For example, it is established for the classical $XY$ model
\cite{KT}, the eight-vertex model solved by Baxter \cite{Baxter}, or for
several two-dimensional Ising models with competing nearest-neighbor (nn) and
next-nearest-neighbor (nnn) interactions, see Refs.
[\onlinecite{Kadanoff71,Barber79,Minami94,Liebmann}] and more references
therein. More examples and discussions on the equivalence between the
two-dimensional classical spin models (Ising, Potts, Ashkin-Teller, $XY$) and
(1+1) quantum models (quantum spin chains, Luttinger, Gaussian), their critical
properties and (non)universality can be found in Refs.
[\onlinecite{Nijs81,Black81}].

From the renormalization group (RG) point of view the non-universality of the
PCA implies that the RG flow for the equivalent equilibrium spin model
\cite{Rujan87,Georges89} would have a manifold of fixed points instead of a
unique fixed point. A paradigmatic example of such behavior is the
Kosterlitz-Thouless picture of RG flow for the $XY$ model, \cite{KT} which
appears also in many (1+1) quantum models. \cite{Giamarchi04} According to
Kadanoff and Wegner, non-universality can be traced back to the presence of an
RG marginal operator. \cite{Kadanoff71}

\begin{figure}[htp]
\centering
\includegraphics[height=18cm]{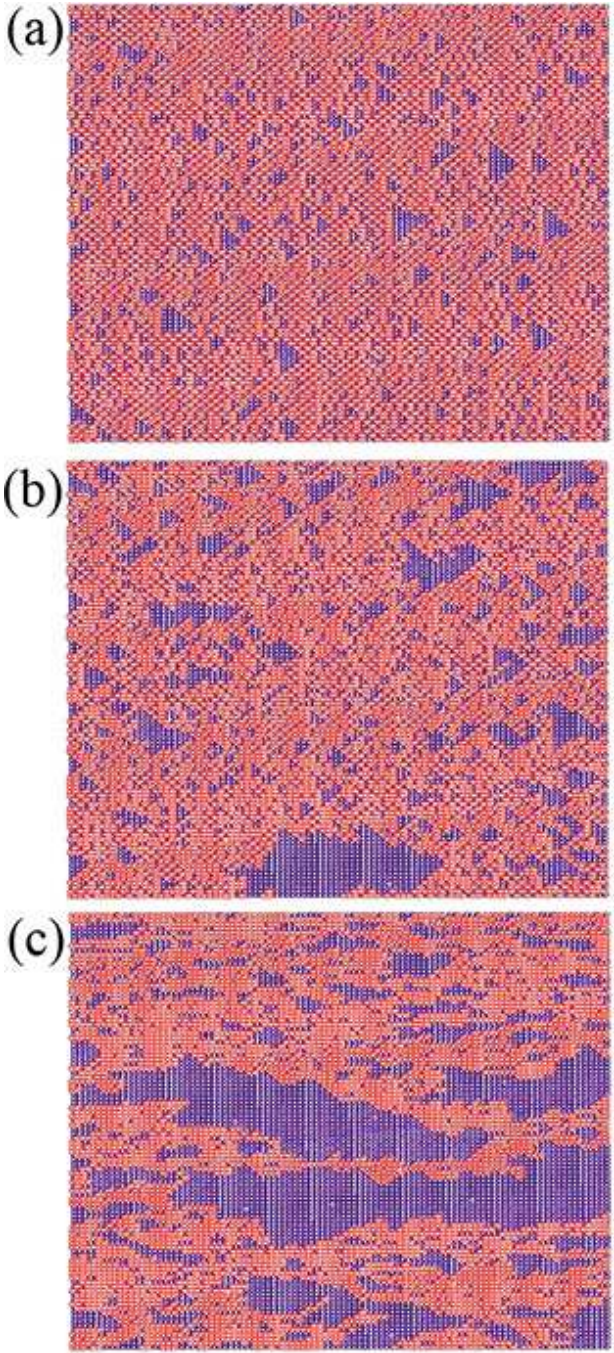}
\caption{(Color online) (a) - $p$ =0.2, $q$ = 0.89, (b) - $p$ =0.5,
$q$ = 0.66, (c) - $p$ =0.8, $q$ = 0.32. Blue sites are empty. The
time direction is horizontal.}\label{Fig.3}
\end{figure}

To corroborate our conjecture of non-universality we point out that
the structure of the active state in this model undergoes
considerable qualitative changes under $p$ and $q$ variations at
constant $n_1$. The space-time patterns of the process which has
reached a steady state with $n_1 \approx 0.5$ are presented in Fig.
\ref{Fig.3} for different $p$ and $q$. The evolution of 100 sites
chosen out of 10000 is shown there during 100 time steps in the
steady state.

\begin{figure}[htp]
\centering
\includegraphics[height=12.0cm]{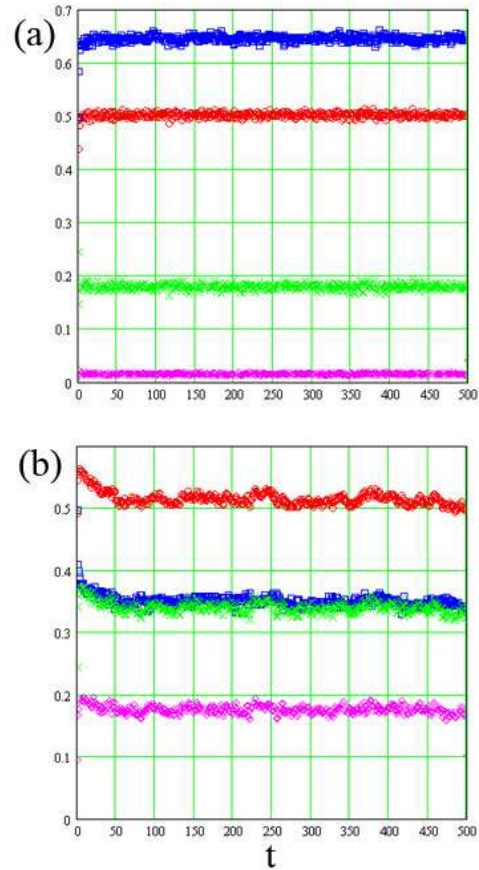}
\caption{(Color online) Time dependence of $n_1$ ($\circ$ red), $n_c$
($\square$ blue), $n_{11}$ ($\times$ green), and $n_{111,1}$ ($\diamond$
magenta); (a) $p$ = 0.2, $q$ = 0.89, (b) $p$ = 0.8, $q$ = 0.32.}\label{Fig.4}
\end{figure}

The pattern difference, quite distinct visually, can be assessed numerically.
It appears that along with $n_1$, the concentration of clusters of adjacent
active and inactive sites in the steady states
\begin{equation}
\label{nc}
    n_c  \equiv  \frac{1}{N}  \sum\limits_{i = 1}^N {\delta \left( {S_i  + S_{i + 1} ,1} \right)}
\end{equation}
stays almost constant with negligible fluctuations, see
Fig.\ref{Fig.4}. In the above equation and throughout, we use
$\delta(m,n)$ as a notation for the conventional Kronecker delta.

This parameter, as well as $n_1$, is almost the same in the different trials.
The values of $n_c$ for three steady states in Fig. \ref{Fig.3} are presented
in Table \ref{tab:table3}. The concentration of active adjacent pairs (11) in
steady configurations $n_{11}$ is also shown there with
\begin{equation}
\label{n11}
    n_{11}  \equiv \frac{1}{N}  \sum\limits_{i = 1}^N {\delta \left( {S_i  + S_{i + 1} ,2} \right)}
\end{equation}
Note that an exact relation
\begin{equation}
\label{n1c}
    n_{11}  = n_1  - n_c /2
\end{equation}
holds. We have also calculated the average values of the supposed marginal
operator
\begin{equation}
\label{n111}
    n_{111,1}  \equiv  \frac{1}{N} \sum\limits_{i = 1}^N {S_{i,t}S_{i + 1,t}S_{i-1,t}S_{i,t+1}}
\end{equation}
This four-spin operator is present in the Hamiltonian of the corresponding
equilibrium spin model and it can be responsible for the non-universality in
($1+1$) dimensions. The average  $n_{111,1}$ is also nearly constant in steady
states (see Fig. \ref{Fig.4}), trial independent and varies with $p$ and $q$,
as shown in  Table \ref{tab:table3}. Its value defines the structural
characteristics of emergent active state and, probably, their indices. Our data
also show that $n_{11}$ (and $n_c$) is proportional to $n_1$ near the
transition points.
\begin{table}
\caption{\label{tab:table3}Parameters of three steady state configurations
shown in Fig. \ref{Fig.3}. }
\begin{ruledtabular}
\begin{tabular}{cccc}
$(p, q)$ & (0.2, 0.89) & (0.5, 0.66) & (0.8, 0.32) \\
$n_1$ & 0.499 & 0.506 & 0.501 \\
$n_c$ & 0.640 & 0.468 & 0.354 \\
$n_{11}$ & 0.179 & 0.272 & 0.324 \\
$n_{111,1}$ & 0.016 & 0.072 & 0.179 \\
\end{tabular}
\end{ruledtabular}
\end{table}

\textit{Summary \& Discussion:} The structural characteristics of
steady states with nearly the same $n_1$ are considerably different,
which is not only seen quite clearly from Fig. \ref{Fig.3}, but it
manifests quantitatively via variations of numbers of clusters per
site $n_c$ (\ref{nc}) and the four-spin operator $n_{111,1}$
(\ref{n111}). Along with the proportionality of $n_{11}$ (and $n_c$)
to $n_1$ this suggests a multi-component order parameter. The latter
violates one of the requirements of the Janssen-Grassberger
hypothesis for a model to belong to the DP universality class
\cite{Hinrichsen06}. We conjecture the non-universality in this
model, most likely due to the marginal four-spin term of the
Hamiltonian [cf. Eq.~(\ref{n111})],  which  manifests itself in
variations of the critical index $\beta$.

However to confirm that this PCA does exhibit the non-universal
critical behavior, the numerical values of other critical indices
are needed. Although it is a rather difficult task for the present
model to get them due to strong fluctuations which other quantities
of interest such as correlation lengths exhibit near the transition
point. Strong fluctuations also hinder determination of the index
$\alpha= 2-\nu_ \| -\nu_\bot$ \cite{Hinrichsen00} from the
energy-like behavior of $n_{111,1} \sim (q-q_c)^{1-\alpha}$.

From the analytical side, the non-universality can be probed by
analysis of the marginal perturbation around an  \textit{integrable}
fixed point. For instance, the non-universality  of the eight-vertex
model or the Ising with competing nn and nnn couplings, can be
demonstrated by the RG analysis of the flow generated by the nn or
four-spin interactions which couple two independent (integrable)
Ising models. \cite{Barber79,Minami94,Nijs81,Black81} In the present
case the RG analysis of non-universality, i.e., of the marginal
perturbations, is more difficult problem, since even the (1+1) DP
fixed point \textit{presumably} controlling the universality, is not
integrable. \cite{Hinrichsen06} Note that the $\epsilon$-expansions
around the DP upper critical dimension $d_c=4$ are not reliable for
our case of $d=1$. \cite{RFT} Clearly, an RG study of the field
theory  corresponding to our model is warranted. It can shed more
light on the the critical properties of the model. In particular,
the field theory formulation of the present PCA model could help to
answer a natural question of how this PCA is distinct from others
known to belong to the DP universality class. \cite{Hinrichsen00} We
plan to address all these issues in the forthcoming paper.
\begin{acknowledgments}
We acknowledge financial support from the Natural Science and
Engineering Research Council of Canada (NSERC) and the Laurentian
University Research Fund (LURF). We thank V. Oudovenko for useful
discussions and his help with the numerical calculations on the
earlier version of this project. P.N.T. thanks Laurentian
University, where the initial stage of the work was done, for
hospitality.
\end{acknowledgments}

%
%
\end{document}